\documentclass{elsarticle}
\usepackage{amsmath,amssymb,amsfonts}
\usepackage{algorithm}
\usepackage{algorithmic}
\usepackage{graphicx}
\usepackage{textcomp}
\usepackage{pdflscape}
\usepackage{makecell}
\usepackage{multirow}
\usepackage{glossaries}
\usepackage{xcolor}

\newacronym{tpm}{TPM}{tree parity machine}
\newacronym{vvtpm}{VVTPM}{vector-valued tree parity machine}
\newacronym{cvtpm}{CVTPM}{complex-valued Tree Parity Machine}
\newacronym{nbtpm}{NBTPM}{non-binary tree parity machine}
\newacronym{ann}{ANN}{artificial neural network}

\newcommand{\etal}{\textit{et al.}}
\biboptions{numbers,sort&compress}
    
\begin{document}
\begin{frontmatter}

\title{Weight Equalization Algorithm for \\ Tree Parity Machines
}
\tnotetext[t1]{Research project partially supported by the program "Excellence initiative -- research university" for the University of Krakow.}


\author[1]{Miłosz Stypiński}
\author[1]{Marcin Niemiec}
\affiliation[1]{organization={Institute of Telecommunications, AGH University of Krakow},
            addressline={Mickiewicza 30},
            postcode={30-059},
            city={Krakow},
            country={Poland}}

\begin{abstract}
Key agreement plays a crucial role in ensuring secure communication in public networks. Although algorithms developed many years ago are still being used, the emergence of quantum computing has prompted the search for new solutions. Tree parity machines have been put forward as a potential solution. However, they possess inherent shortcomings, one of which is the uneven distribution of values in the secured key obtained after the key agreement process, especially when on-binary vectors are used during the synchronization process. This paper introduces a novel algorithm designed to address this issue. The results demonstrate a substantial enhancement in the quality of the secured key obtained.
\end{abstract}

\end{frontmatter}

\section{Introduction}

Secure key agreement refers to a category of protocols in which two or more participants employ established algorithms to mutually generate a cryptographic key. Despite sharing all the necessary information for key agreement over a publicly insecure channel, potential attackers are incapable of deducing any information about the key. 

The Diffie-Hellman algorithm~\cite{difhel} and its elliptic curve variant~\cite{ecdh} are among the most widely used key exchange algorithms. Both of them are backed by algebraic number theory problems, which are supposed to be difficult to revert. However, successful implementation of Shor's algorithm of sufficiently powerful quantum computers would be capable of breaking the security of widely used key exchanged algorithms~\cite{shor}. This issue has become sufficient to create a list of four NIST-approved quantum-proof algorithms~\cite{Moody2023}.

One of the alternative quantum proof algorithms for key distribution and agreement is mutual learning using specific kind of neural networks -- \gls{tpm}. \Glspl{tpm} have been extensively studied \cite{TPM_1,TPM_2,Kanter_2002,Andreas_Ruttor_2004,PHD,CVTPM,VVTPM,NBTPM,Sarkar2019,Niemiec2019,GOMEZ2017430,stypinski_3}. In~\cite{TPM_1,TPM_2,Kanter_2002,Andreas_Ruttor_2004} the authors introduced an innovative key agreement protocol employing \glspl{ann}. Through mutual learning and learning rule definition they have achieved synchronization of \glspl{tpm}. Additionally, the authors have demonstrated that \gls{tpm} synchronization finishes in finite time. The systematic presentation of this work can be found in~\cite{PHD}.

Several enhancements have been suggested for \glspl{tpm}, including Dong et al.'s proposal to use complex values instead of binary values during the learning process~\cite{CVTPM}. This idea was further generalized in~\cite{VVTPM}, where the authors introduced the use of binary vector values as inputs. In~\cite{NBTPM} the authors have proposed a usage of integer inputs instead of binary. These improvements are named, accordingly, the \gls{cvtpm}, \gls{vvtpm}, and \gls{nbtpm}. This paper focuses on exploring the latter.

\Glspl{tpm} have applications in numerous fields. Sarkar \etal have proposed the usage of neural network synchronization in wireless systems \cite{Sarkar2019}. In~\cite{Niemiec2019} \glspl{tpm} are used as key reconciliation mechanism in quantum key distribution networks. Another application is described in~\cite{GOMEZ2017430} where \gls{tpm} was responsible for key establishment on microprocessors. Mutual learning, as described in~\cite{stypinski_3}, could also find applications in the field of smart grids.

The aim of this article is to propose and evaluate a novel algorithm for \gls{tpm} weight equalization. The problem is significant in networking environment as we are currently in the search of alternatives to key agreement protocols base on factorization problem. The use of \glspl{nbtpm} and proposed algorithms significantly improves the communication security features and renders \glspl{tpm} as an alternative to currently used key agreement protocols. Furthermore, \glspl{nbtpm} result in smaller key agreement protocol overhead compared to standard \gls{tpm} because lesser number of mutual learning iterations are required to successfully generate secure key. 

The purpose of this article is to present and assess a novel algorithm for weight equalization in \gls{tpm}. The problem is significant in networking environments, particularly as there is a need for alternatives to key agreement protocols based on factorization problems. Incorporating \glspl{nbtpm} and the proposed algorithms brings notable enhancements in communication security, positioning \gls{tpm} as a viable substitute for current key agreement protocols. Moreover, \gls{nbtpm} leads to a reduced key agreement protocol overhead compared to standard \gls{tpm}, as it requires a lesser number of mutual learning iterations for the successful generation of a secure key.
\section{Tree Parity Machine}
The original \Gls{tpm} is a two-layered, binary input, binary output \gls{ann}~\cite{TPM_1,TPM_2,Kanter_2002,Andreas_Ruttor_2004}. The hidden layer consists of $K$ neurons with $N$ inputs per neuron. With every input, there is a weight associated with it that is an integer and is constrained within the range from $-L$ to $L$ and where $L$ is a natural number greater than $0$. Similarly, in the feed-forward neural network, in \gls{tpm} all the outputs of the layer are connected to the neurons of the succeeding layer. \Gls{nbtpm} alters the values that the input vectors can take and allows them to take values from the range of $-M$ to $M$, where $M$ is a natural number greater than $0$~\cite{NBTPM}. It is worth noting that \gls{nbtpm} is a generalization of the typical \gls{tpm}, and for $M = 1$ there is no difference between these two. To describe a specific \gls{nbtpm} one needs the quadruple of parameters $K$, $L$, $M$, $N$ which, respectively, denote the number of hidden neurons, range of weight values, variability of input vectors and number of inputs per neuron in hidden layer. Due to the requirement of integer weights in the described model, backpropagation is not applicable in \glspl{tpm}. Therefore, other learning methods are needed, which will be described later in this section.

To maintain the binary nature of \glspl{tpm} the activation function of neurons in the hidden layer is a function returning either $-1$ or $1$. Specifically, this is an altered signum function ($f_{signum}$) that does not produce a $0$ result. Instead, it returns $-1$ or $1$, depending on whether it represents the communication of the sender or the recipient. The formula to calculate the output of neurons is presented in \eqref{neuron_output}, where $y_k$ denotes the output of the k-th neuron, $x_{kn}$ denotes the input of the $n$-th neuron, and $w_{kn}$ denotes the weight associated with the input.

\begin{equation}
  y_k = f_{signum} (\sum_{n=1}^{N}x_{kn} \cdot w_{kn} ) \label{neuron_output}
\end{equation}

\begin{figure}[htbp]
  \centerline{
    \includegraphics[width=0.8\columnwidth]{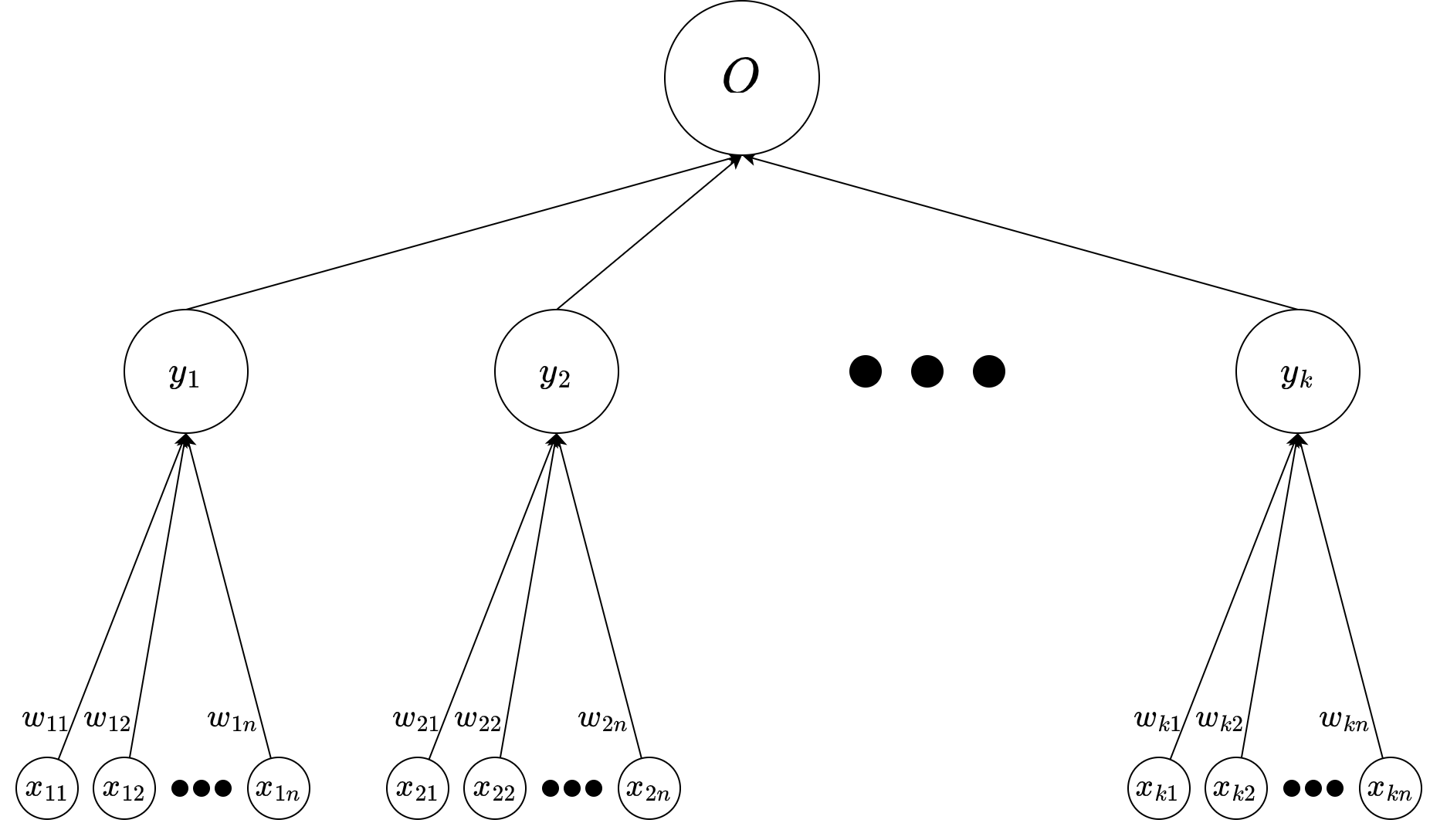}
  }
  \caption{Architecture of the \acrlong{tpm}}
  \label{tpm_architecture}
\end{figure}

\Gls{tpm} output is defined by $O = \prod_{k}^{K} y_k$ where $y_k$ denotes the output from the $k$-th hidden neuron. Since $y_k$ gives a binary value, the final output is also binary.
The described \gls{tpm} architecture is presented in Figure \ref{tpm_architecture}.

Mutual learning, in the context of \gls{tpm}, refers to a specialized key agreement protocol. It involves two neural networks that execute a predefined algorithm to establish synchronization between them. Synchronization is an iterative process and results in both neural networks having exactly the same weights, which can be used as a shared secret in further cryptographic operations. The learning algorithm is responsible for updating the weights. During each successful iteration step, the weights are updated in a manner that draws both TPMs closer to each other. Key agreement protocols assume that there is a public channel which may be eavesdropped on. The steps of mutual learning are summarized below~\cite{PHD}.
\begin{enumerate}
    \item Key agreement parties agree on an exact value for \gls{tpm} parameters ($K$, $L$, $M$, $N$) and initialize their own copy of \gls{tpm} with randomly chosen weights.
    \item One of the participants generates a random input vector $X$ the values of which are bound by $M$ and shares it with the other.
    \item Both parties calculate the output of their \gls{tpm} and share it publicly. In case the outputs match, they employ one of the following learning rules:
        \begin{itemize}
          \item	\textit{Hebbian learning rule}
                \begin{equation}
                  w_{kn}(t+1) = w_{kn}(t) + O(t)x_{kn}(t)\Theta(y_k(t), O(t)),
                  \label{eq_hebbian_lr}
                \end{equation}
          \item	\textit{Anti-Hebbian learning rule}
                \begin{equation}
                  w_{kn}(t+1) = w_{kn}(t) - O(t)x_{kn}(t)\Theta(y_k(t), O(t)),
                  \label{eq_antihebbian_lr}
                \end{equation}
          \item	\textit{random walk learning rule}
                \begin{equation}
                  w_{kn}(t+1) = w_{kn}(t) + x_{kn}(t)\Theta(y_k(t), O(t)),
                  \label{eq_random_walk_lr}
                \end{equation}  
            where $t$ denotes the t-th iteration step and $\Theta$ is a function with returns $1$ if all its arguments are equal.
        \end{itemize}
    \item Steps 2-3 are repeated until full synchronization is achieved.
\end{enumerate}

Once mutual learning is completed, both \glspl{tpm} are synchronized and have their corresponding weights equal. Both parties are able to use the distilled weights as their shared secret for further cryptographic purposes. The secret length is variable and depends on the size of the \gls{tpm}. Assuming an equal distribution of values in the weight vector $W$ the secret length would be equal to $K\cdot~N\cdot~log_2(2L + 1)$. However, studies show that the distribution differs from the uniform distribution~\cite{NBTPM, PHD}, and therefore the formula needs to be updated to take this imperfection into account. In~\cite{NBTPM} the authors defined a formula that meets these conditions and presented it in~\eqref{eq_key_length}, where $K$ and $N$ are the parameters of \gls{tpm} and $E(W)$ is an entropy of the weight vector $W$.
\begin{equation}
  len_{secret} = K \cdot N \cdot E(W) = K \cdot N \cdot (-\sum_{l=-L}^{L}p_llog_2p_l)
  \label{eq_key_length}
\end{equation}

\section{Weight equalization}

\gls{nbtpm} shortens the mutual learning process that results in improved security features of the key agreement protocol. The cost of this improvement is a more uneven distribution of the weight vector values. The values $w_{kn}$, where $|w_{kn}|= L$ occur more frequently than other values. This phenomenon is known as the Extrema Values Effect and is described in detail in~\cite{NBTPM}. A similar phenomenon occurs in standard \gls{tpm}, albeit with a less pronounced effect. This paper proposes a novel algorithm in terms of \gls{tpm}, later called weight equalization, that equalizes the probability of the occurrence of frequent values in the weight vector. The algorithm is inspired by histogram equalization~\cite{histogram_eq}.

The weight equalization algorithm comprises the equalization, dropout, and substitution phase. 
The equalization phase is responsible for replacing more commonly occurring values with those that occur less frequently. To execute this algorithm, a weight vector $W$ is required that has been derived from synchronized \glspl{tpm}. Additionally, the algorithm takes a set of parameters $(K, L, N)$ that define the \gls{tpm} size as the input. During the equalization phase, the algorithm goes through each element of the weight vector $W$. If, during current iteration, the current weight $w_{kn}$ is the most frequently occurring one, it is exchanged with the least frequent value encountered so far in the vector. To identify the least frequent value among the already processed values, the algorithm keeps a count of how many times each value has appeared, using a vector called $F$ as a cache. This weight equalization process is outlined in Algorithm~\ref{algorithm_1}. It is important to note that in this algorithm, the functions argmax and argmin return sets of indices where the maximum and minimum values occur, respectively. Furthermore, the algorithm updates the values in place, so the weight vector must be mutable.

\begin{algorithm}[H]
    \caption{Equalization phase}
    \begin{algorithmic}
    \renewcommand{\algorithmicrequire}{\textbf{Input:}}
    \renewcommand{\algorithmicensure}{\textbf{Output:}}
        \REQUIRE $W \gets [w_{11}, \cdots, w_{1n}, w_{21}, \cdots, w_{kn}]$, $K$, $N$, $L$
        \ENSURE $W$
        \STATE $F \gets [f_{-L}, \cdots, f_L] = [0, \cdots, 0]$
        \STATE $k \gets 1$
        \WHILE {$k \leq K$}
            \STATE $n \gets 1$
            \WHILE {$n \leq N$}
                \IF{$W_{kn} \in argmax\,F$}
                    \STATE $w_{kn} \gets min(argmin\,F)$
                \ENDIF
                \STATE $F_{w_{kn}} \gets F_{w_{kn}} + 1$
                \STATE $n \gets n+1$
            \ENDWHILE
            \STATE $k \gets k+1$
        \ENDWHILE 
    \end{algorithmic}
    \label{algorithm_1}
\end{algorithm}

The dropout phase is responsible for improving the quality of the distilled key. Once the equalization is complete, the key length is longer than the limit created by equation~\eqref{eq_key_length}. To address this issue the dropout phase was incorporated into the algorithm. Algorithm~\ref{algorithm_2} keeps track of the current secret length, and if the weight exceeds the theoretical limit at some point, it is dropped instead.

\begin{algorithm}[H]
    \caption{Dropout phase}
    \begin{algorithmic}
    \renewcommand{\algorithmicrequire}{\textbf{Input:}}
    \renewcommand{\algorithmicensure}{\textbf{Output:}}
        \REQUIRE $K$, $N$, $L$, $W$ before equalization
        \ENSURE $W'$
        \STATE $E(W) = -\sum_{l=-L}^{L}p_llog_2p_l$
        \STATE $len_{current} \gets 0 $
        \STATE $W'$, which is a variable length vector with elements
        \STATE $k \gets 1$
        \WHILE {$k \leq K$}
            \STATE $n \gets 1$
            \WHILE {$n \leq N$}
                \IF{$len_{current} < ((k-1) \cdot N+n) \cdot E(W)$}
                    \STATE $W'.append(w_{kn})$
                    \STATE $len_{current} \gets len_{current} + log_2(2L+1)$
                \ENDIF
                \STATE $n \gets n+1$
            \ENDWHILE
            \STATE $k \gets k+1$
        \ENDWHILE 
    \end{algorithmic}
    \label{algorithm_2}
\end{algorithm}

The final stage of the algorithm involves substitution. The objective in this phase is to further enhance the randomness of the secret and propagate single errors to affect the entire secret. The most suitable tools for achieving this goal are cryptographic hash functions, which, thanks to the avalanche effect, exhibit substantial changes in their output even when there are minor variations in the input. Furthermore, these functions are difficult to reverse, offering an additional advantage. During this step, the secret is divided into blocks, each with a length equal to that of the hash function's output. Subsequently, each block is replaced with the output of the hash function for that specific block. If the last block happens to be shorter than the hash function's output, it is simply omitted.

Both of the mutual learning participants execute the presented algorithm to improve their secret key quality. The algorithm is deterministic, hence both parties will always obtain the same secret once the algorithm is executed.

\section{Evaluation}

The algorithm's performance assessment involves two main aspects. First, it involves comparing the probabilities of attaining specific weights in the weight vector both before and after executing the algorithm. Second, the evaluation also includes a comparison of the results of the NIST Test Suite~\cite{NIST} for the weight vector after representing it in binary form, both before and after the application of the weight equalization algorithm. To obtain the weight vectors, mutual learning has been performed using \gls{nbtpm} with following parameters: $K=3$, $L=8$, $M=\{1;3;5\}$, and $N=60$. Mutual learning was performed using the Hebbian learning rule. The \Glspl{tpm} were synchronized $1000$ times using a dedicated simulation framework. The selected hash function for the substitution phase was SHA-256.

NIST Special Publication 800-22 defines a set of tests that aims to assess the quality of random and pseudo-random number generators for cryptographic applications~\cite{NIST}. The following is the list of tests with a brief description and selected parameters. If the test parameters are not provided, the default values defined in \cite{NIST} are used. 

\begin{enumerate}
    \item \textit {Frequency (Monobit) Test} -- aims to check the ratio of zeroes and ones in a random sequence.
    \item \textit {Frequency Test within a Block} -- evaluates proportions of ones in $M$-bit block. The block size used in this paper is equal to $128$ bits.
    \item \textit {Runs Test} -- analyzes whether the oscillation between ones and zeros stays within the boundary for random sequence.
    \item \textit {Test for the Longest Run of Ones in a Block} -- examines the same bit sequence in an $M$-length block.
    \item \textit {Binary Matrix Rank Test} -- checks the linear dependence between the subsequences of the original random values.
    \item \textit {Discrete Fourier Transform (Spectral) Test} -- looks for anomalies in the Discrete Fourier Transform of the random sequence
    \item \textit {Non-overlapping Template Matching Test} -- checks the frequency of occurrences of the $M$-bit sequence defined beforehand.
    \item \textit {Overlapping Template Matching Test} -- the aim of this test is the same as the previous one; however, if it finds a sequence it only moves the search window only by one bit instead of $M$ bits.
    \item \textit {Maurer’s Test} -- evaluates whether the examined sequence is compressible using a lossless algorithm.
    \item \textit {Linear Complexity Test } -- evaluates the length of a linear feedback shift register. The block used for this test is equal to $500$ bits.
    \item \textit {Serial Test} -- tests the frequency of all $M$-bit permutations in the examined sequence.
    \item \textit {Approximate Entropy Test} -- evaluates the occurrence rate of every conceivable $M$-bit pattern.
    \item \textit {Cumulative Sums (Cusum) Test } -- assesses whether the cumulative sums of a random binary string are within an acceptable range for a particular random sequence. This assessment is carried out both forwards and backwards.
    \item \textit {Random Excursions Test} -- analyses how many times the specific state has been visited during the cumulative sums random walk performed on the sequence. Before running the test, $0$ in a random value is swapped for the value $-1$. The results are presented for states ranging from $-4$ to $4$ omitting $0$.
    \item \textit {Random Excursions Variant Test} -- test methodology is the same as the previous one; however, the states consider the range from $-9$ to $9$ without 0.
\end{enumerate}

\begin{figure}[htbp]
  \centerline{
    \includegraphics[width=0.9\columnwidth]{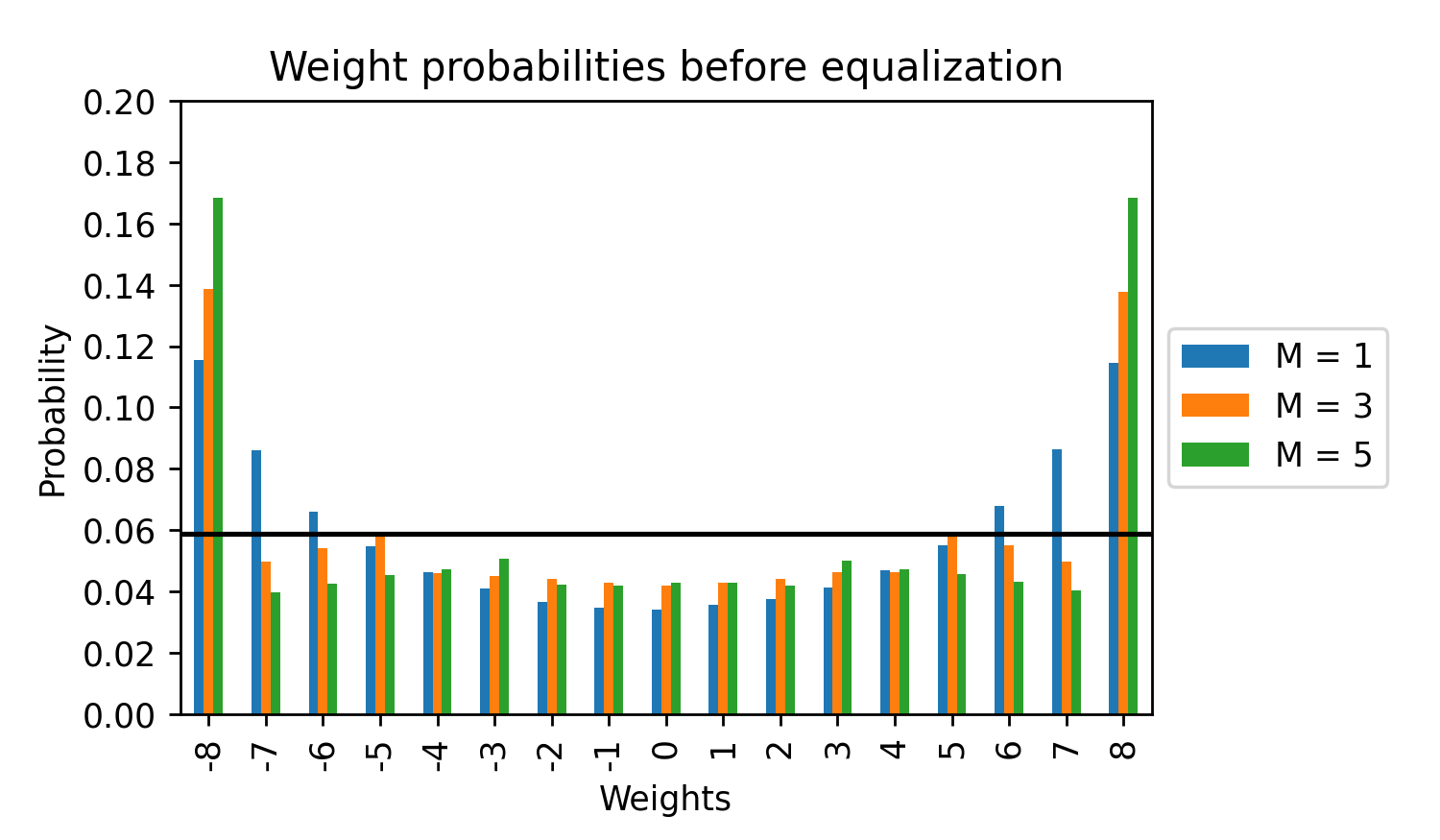}
  }
  \caption{Weight probabilities before equalization}
  \label{chart_1}
\end{figure}

\begin{figure}[htbp]
  \centerline{
    \includegraphics[width=0.9\columnwidth]{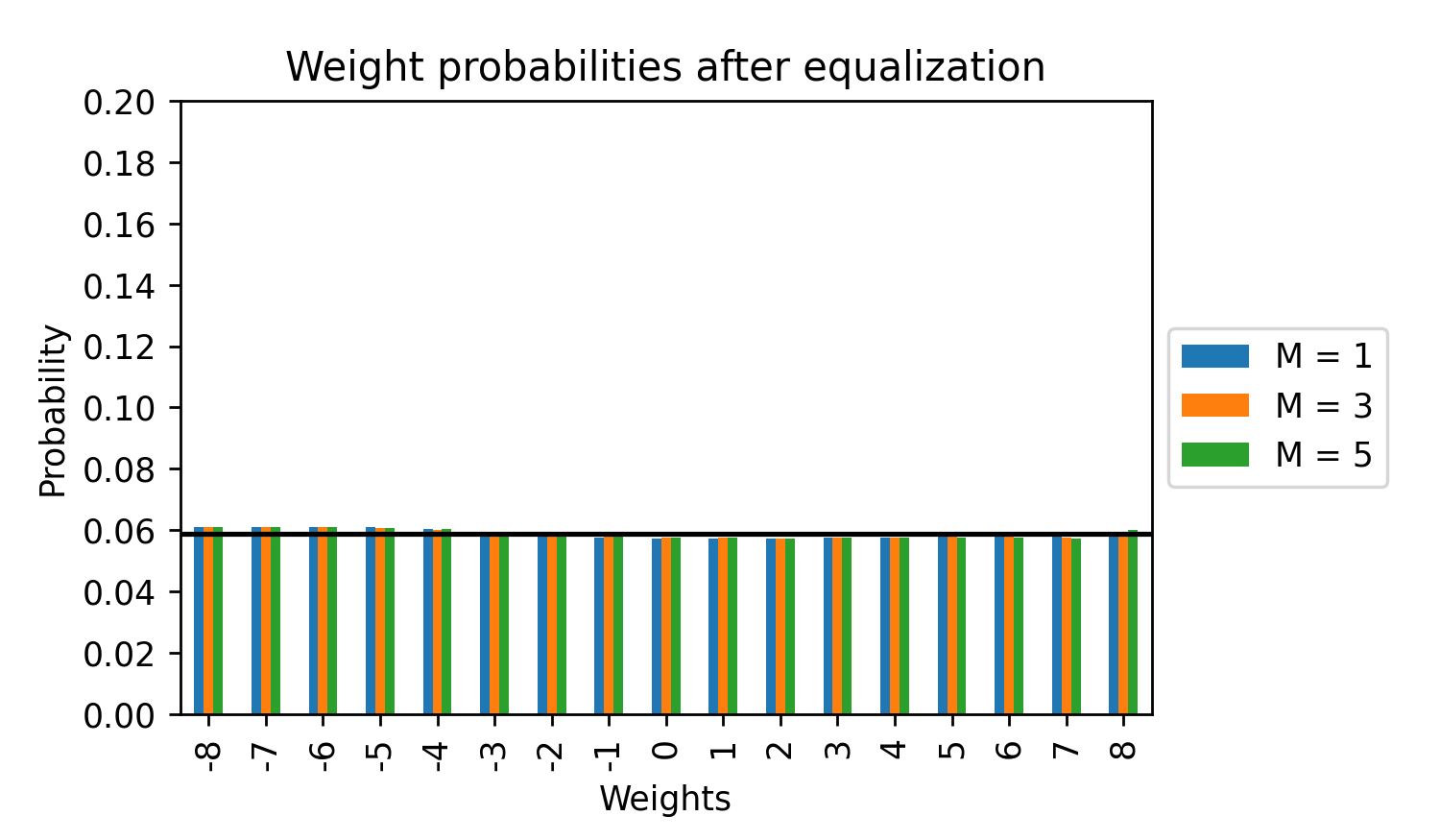}
  }
  \caption{Weight probabilities after equalization}
  \label{chart_2}
\end{figure}

Firstly, the probability of the presence of a specific value in the weight vector is examined. The probabilities of weight occurrence are not equal. With the increase of $M$ the distribution becomes more dominated by limit values. While for $M = 1$ the probability of $w_{kn} \in \{-L; L\}$ is equal to $0.23$, for $M = 3$ and $M=5$ the same probability is equal to $0.28$ and $0.34$, respectively. This is an example of the Extrema Values Effect phenomenon. The probabilities for all weights before running the equalization algorithm are detailed in Figure~\ref{chart_1}. The black horizontal line depicts the ideal uniform distribution $y = \frac{1}{2L + 1}$

Following equalization, the probabilities follow an almost uniform distribution. The greatest difference from ideal uniform distribution is $0.0023$, which is equal to $3.89\% $ of $\frac{1}{2L + 1}$. For comparison, the weights following equalization are presented in Figure~\ref{chart_2}.

Another evaluation method consists of testing the weight vector against the NIST testing suite. The tests return on or more $P$-values which are the determinants of the probability that the tested sequence is fully random. If the $P$-value is equal to $1$ then the tested sequence is fully random. On the other hand, if the $P$-value is equal to $0$, then the tested sequence is generated in a deterministic manner. The test is successful when the returned $P$-value is greater than $0.01$. Tests have been run on the weight vector obtained from the aforementioned simulations. The weights have been encoded using little-endian encoding. Before applying the equalization algorithm, the values of $0$ have been removed from the weight vector to facilitate encoding.

The tests demonstrate that direct usage of \gls{tpm} and \gls{nbtpm} produces a secret key that cannot be used for security purposes. Only two tests passed for all \gls{tpm} variants before using the equalization algorithm: the \textit{Binary Matrix Rank Test} and \textit{Linear Complexity Test}. On the other hand, the weight equalization algorithm significantly improves the overall results of the test suite on the obtained key. The only tests with produce allegedly poorer results for specific scenarios are the \textit{Random Excursions Test} and \textit{Random Excursions Test Variant} for some states. However, the sequence produced is pseudo-random and could have been drawn in such a way that gave poorer results in this case. Moreover, the result of the \textit{Approximate Entropy Test} for $M = 5$ was close to a failure. This conclusion follows studies conducted in~\cite{NBTPM} where the authors stated that the values of $M$ approaching the values of $L$ results in a less secure \gls{nbtpm} variant.

\begin{table}

    \centering
    \rotatebox{90}{
    \begin{minipage}{0.9\paperwidth}
    \caption{NIST Test Suite results}
    
    \label{table_results}
    \begin{tabular}{|c|c|c|c|c|c|c|}
                \cline{2-7}
             \multicolumn{1}{c|}{} &  \multicolumn{6}{c|}{$P$-values} \\
            \hline
            \multirow{2}{*}{Test} & \multicolumn{2}{c|}{$M = 1$}  & \multicolumn{2}{c|}{$M = 3$} & \multicolumn{2}{c|}{$M = 5$} \\
                \cline{2-7}
            & b. e. &  a. e. & b. e. &  a. e. & b. e. & a. e.\\
            \hline
1)							 & 0.0 & 0.961 & 0.0 & 0.665 & 0.0 & 0.397 \\
2)						 & 0.0 & 0.66 & 0.0 & 0.011 & 0.0 & 0.734 \\
3)									 & 0.0 & 0.063 & 0.0 & 0.686 & 0.0 & 0.155 \\
4)				 & 0.0 & 0.726 & 0.0 & 0.073 & 0.0 & 0.394 \\
5)					 & 0.461 & 0.587 & 0.791 & 0.953 & 0.104 & 0.491 \\
6)	 & 0.0 & 0.171 & 0.0 & 0.623 & 0.0 & 0.517 \\
7)		 & 0.0 & 0.052 & 0.0 & 0.369 & 0.0 & 0.668 \\
8) 			 & 0.0 & 0.443 & 0.0 & 0.268 & 0.0 & 0.462 \\
9)				 & 0.0 & 0.518 & 0.0 & 0.135 & 0.0 & 0.903 \\
10)					 & 0.126 & 0.813 & 0.294 & 0.337 & 0.535 & 1.0 \\
11)								 & (0.0, 0.711) & (0.522, 0.788) & (0.0, 0.028) & (0.717, 0.452) & (0.0, 0.0) & (0.713, 0.667) \\
12)					 & 0.0 & 0.3 & 0.0 & 0.477 & 0.0 & 0.025 \\
13) (Fwd)				 & 0.0 & 0.676 & 0.0 & 0.812 & 0.0 & 0.746 \\
13) (Bwd)				 & 0.0 & 0.63 & 0.0 & 0.758 & 0.0 & 0.583 \\
 \hline
14) & 
\makecell{0.845, 0.278, \\ 0.89, 0.915, \\
0.626, 0.24, \\ 0.082, 0.25} & 
\makecell{0.982, 0.402, \\  0.113, 0.56, \\
0.685, 0.933, \\ 0.721, 0.109} & 
\makecell{0.014, 0.0, \\  0.341, 0.021, \\
0.514, 0.883, \\  0.211, 0.053} & 
\makecell{0.418, 0.351, \\  0.631, 0.638, \\
0.93, 0.642, \\  0.423, 0.468} & 
\makecell{0.926, 0.932, \\  0.232, 0.579, \\
0.326, 0.877, \\  0.046, 0.021} & 
\makecell{0.394, 0.216, \\  0.085, 0.396, \\
0.028, 0.346, \\ 0.805, 0.478} \\
\hline
15) & 
\makecell{0.42, 0.374, 0.357,\\ 0.414, 0.512, 0.71, \\
0.783, 0.722, 0.538, \\ 0.806, 0.67, 0.545, \\
0.816, 0.712, 0.436, \\ 0.357, 0.325, 0.339} & 
\makecell{0.088, 0.104, 0.207, \\ 0.411, 0.537, 0.564, \\
0.331, 0.113, 0.108, \\ 0.745, 0.63, 0.949, \\
0.914, 0.65, 0.569, \\ 0.437, 0.358, 0.481} & 
\makecell{0.903, 0.949, 1.0, \\ 1.0, 0.803, 0.637, \\
1.0, 0.386, 0.317, \\ 0.617, 0.773, 0.655, \\
0.705, 0.677, 0.546, \\ 0.579, 0.606, 0.628} & 
\makecell{1.0, 0.974, 0.967, \\ 0.821, 0.588, 0.755, \\
0.56, 0.236, 0.293, \\ 0.453, 0.174, 0.268, \\
0.684, 0.707, 0.734, \\ 0.934, 0.821, 0.552} & 
\makecell{0.878, 0.87, 0.861, \\ 1.0, 1.0, 0.811, \\
0.671, 0.855, 0.527, \\ 0.752, 0.715, 0.888, \\
0.72, 0.673, 0.634, \\ 0.661, 0.683, 0.701} & 
\makecell{0.53, 0.33, 0.258, \\ 0.228, 0.195, 0.259, \\
0.224, 0.142, 0.141, \\ 0.948, 0.175, 0.103, \\
0.3, 0.539, 0.401, \\ 0.265, 0.308, 0.287} \\
 \hline
    \end{tabular}
    \end{minipage}
    }
\end{table}

The results obtained from the simulations are presented in Table~\ref{table_results}. Abbreviations \textit{b. e.} and \textit{a. e.} in the table header mean the results before and after equalization, respectively. For the \textit{Random Excursions Test} and \textit{Random Excursions Variant Test} results are sorted by states from $-4$ to $4$ and from $-9$ to $9$ respectively.

\section{Summary}
This paper introduces a novel algorithm designed for weight equalization in \glspl{tpm}. The algorithm consists of three phases: equalization, dropout, and substitution.  Its utilization addresses the issue of uneven weight distribution that may arise after synchronization of \glspl{tpm}. Moreover, it effectively mitigates the disadvantages introduced by the use of \gls{nbtpm}, the Extrema Values effect in particular. The paper also investigates the impact of the algorithm on the resulting security key. It examines the probability of specific weight values appearing in the weight vector before and after applying the algorithm. Additionally, the secret key generated is subjected to testing using the NIST randomness test suite. Future research should focus on examining the quality of the key obtained after using the proposed algorithm and determining whether its use facilitates potential attacks on the mutual learning process.

\glspl{tpm} are neural networks that have applications in key synchronization and key reconciliation processes. By utilizing improvements, such as the proposed algorithm, \glspl{tpm} becomes more secure and more competitive against other algorithms.

\bibliography{refs} 
\bibliographystyle{ieeetr}

\end{document}